\documentclass{jfm}
\usepackage{graphicx}
\usepackage{mathrsfs}
\usepackage{epstopdf, epsfig,subfigure}
\usepackage{amsmath}
\usepackage{gensymb}
\usepackage{booktabs}
\usepackage{multirow}

\shorttitle{Prediction of shock wave configurations}
\shortauthor{Yan-Chao Hu, Wen-Feng Zhou, Yan-Guang Yang, Zhi-Gong Tang and Zhao-Hu Qin}

\title{Prediction of shock wave configurations in compression ramp flows}

\author{Yan-Chao Hu\aff{1}
  \corresp{\email{yanchaohu@pku.edu.cn}},
  Wen-Feng Zhou\aff{2},
  Yan-Guang Yang\aff{3}
  \corresp{\email{yangyanguang@cardc.cn}},
  Zhi-Gong Tang\aff{3}
 \and Zhao-Hu Qin}

\affiliation{\aff{1}Hypervelocity Aerodynamics Institute, China Aerodynamics Research and Development Centre, Mianyang 621000, China
\aff{2}State Key Laboratory for Turbulence and Complex Systems and Department of Mechanics,College of Engineering, Peking University, Beijing 100871, China
\aff{3}China Aerodynamics Research and Development Centre, Mianyang 621000, China}

\begin{document}

\maketitle

\begin{abstract}
Here, we provide a theoretical framework revealing that a steady compression ramp flow must have the minimal dissipation of kinetic energy, and can be demonstrated using the least action principle. For a given inflow Mach number $M_{0}$ and ramp angle $\alpha$, the separation angle $\theta_{s}$ manifesting flow system states can be determined based on this theory. Thus, both the shapes of shock wave configurations and pressure peak $p_{peak}$ behind reattachment shock waves are predictable. These theoretical predictions agree excellently with both experimental data and numerical simulations, covering a wide range of $M_{0}$ and $\alpha$. In addition, for a large separation, the theory indicates that $\theta_{s}$ only depends on $M_{0}$ and $\alpha$, but is independent of the Reynolds number $Re$ and wall temperature $T_{w}$. These facts suggest that the proposed theoretical framework can be applied to other flow systems dominated by shock waves, which are ubiquitous in aerospace engineering.
\end{abstract}


\section{Introduction}
\par
\begin{figure}
	\centerline{\includegraphics[width = 0.84\columnwidth]{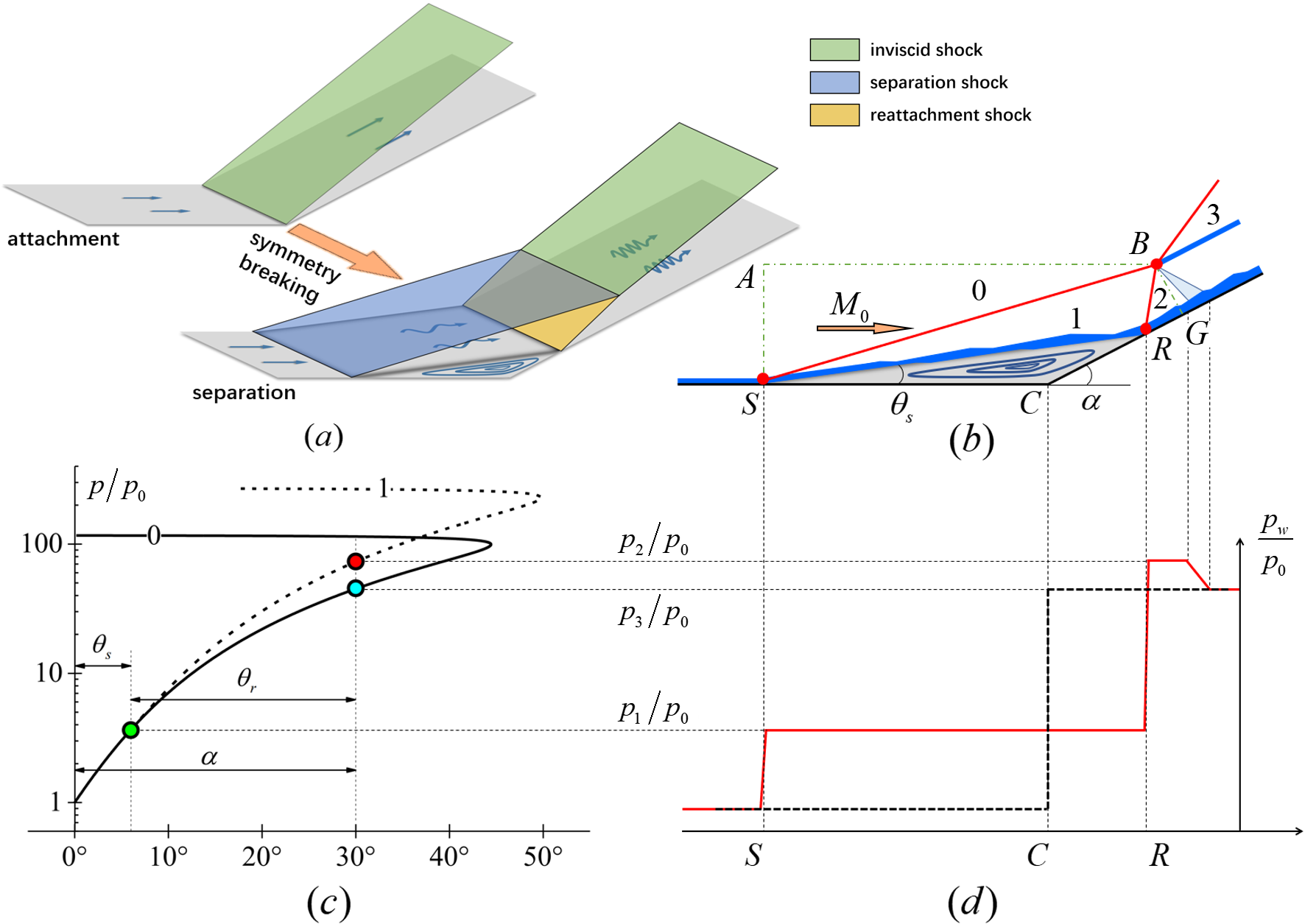}} 
	\caption{(a) Flow patterns of attachment and separation, where inviscid, separation, and reattachment shock waves are green, blue, and orange, respectively; (b) Two types of discontinuity dividing the flow system into four subsystems (0, 1, 2, and 3), where the red and blue parts represent shock waves and shear layers, respectively; (c) Shock polars of this flow system, where polars 0 and 1 correspond to subsystem 0 and 1, respectively. The green, cyan, and red points correspond to $p_{1} / p_{0}$, $p_{3} / p_{0}$ and $p_{2} / p_{0} = p_{peak} / p_{0}$, respectively; (d) Distributions of wall pressure of flow systems, where the red and black lines correspond to the attachment and separation states, respectively.}
	\label{fig:compression-ramp-information}
\end{figure}

 Compression ramp flows are canonical complex flows of interactions between shock waves and boundary layers and are ubiquitous in aerospace engineering (\cite{babinsky2011shock}). As Figure \ref{fig:compression-ramp-information} (a) shows, when a large separation occurs, the flow pattern will change significantly, and such a shock pattern was classified as type VI by \cite{edney1968anomalous}. In this process, a recirculation region, called a `separation bubble', will emerge and induce two new shock waves, i.e., the separation and reattachment shock waves. As a result, spatial distributions of physical quantities, such as the flow pressure $p$, will change. The geometrical features of the separation bubble have a significant contribution that its size (which can be described by the area of the bubble $\Omega_{s}$) and shape (the separation angle $\theta_{s}$) determine the range and intensity of the wall pressure $p_{w}$, respectively, as shown in Figures \ref{fig:compression-ramp-information} (b), (c), and (d). Many numerical (\cite{hung1976numerical,rudy1989validation,olejniczak1998computation,deepak2013computational}) and experimental (\cite{holden1966experimental,Lewis1968Experimental,delery1986shock,mallinson1997interaction}) studies have been conducted and they primarily focused on the size (\cite{burggraf1975asymptotic,rizzetta1978triple,daniels1979laminar,korolev2002once}) and internal details (\cite{smith1991interactive,korolev2002once,neiland2008temperature,shvedchenko2009secondary,gai2019hypersonic}) of the separation bubble. However, only a few of these studies discuss the bubble shape $\theta_{s}$, even though it is a pivotal parameter in determining the flow field.
\par Because the emergence of the separation bubble decreases the homogeneity degree of the flow system (symmetry-breaking), the process from the attachment state ($\theta_{s} = 0$) to the separation state ($\theta_{s} > 0$), shown in Figure \ref{fig:compression-ramp-information} (a), can be analogous to a type of phase transition, where $\theta_{s}$ manifesting the flow states is actually an order parameter --- the concept introduced by \citet {landau1937theory} for phase transition \citep[see][pp. 136-137]{goldenfeld2018lectures}. Therefore, the separation bubble is a `dissipative structure'---the concept proposed by \citet {prigogine1978time} for structures emerging in systems far from equilibrium---and its self-organization process must be governed by the synergic principle of the subsystems (0,1,2, and 3 that are divided by shock waves and shear layers), shown in Figure \ref{fig:compression-ramp-information} (b). From this perspective, this principle is expected to be expressed in an integral form, consisting of the differential governing equation, so that the synergy of this flow system can be described and understood easily. For this purpose, we use the least action principle to establish the equivalent form between the differential and integral scales.
\par In this paper, we demonstrate that the synergic principle of this flow system is the minimal dissipation theorem. Based on this theorem, the separation angle $\theta_{s}$ can be determined; subsequently, the flow patterns are predictable. The proposed theoretical predictions agree excellently with the numerical and experimental results for a wide range of Mach numbers $M_{0}$ and ramp angles $\alpha$. Additionally, the theorem indicates that shapes of shock wave configurations induced by large separations are independent of the Reynolds number $Re$ and the wall temperature $T_{w}$.
\section{The minimal dissipation theorem}
\par In this section, we will demonstrate that the flow system has minimal dissipation. The steady states of the flow system satisfy the following equations:
\begin{align}
& \text{mass equation:}\quad &\nabla \cdot(\rho \boldsymbol{u}) =0 \label{eq:Mass_conservation}\\ 
& \text{momentum equation:}\quad &\rho \boldsymbol{u} \cdot \nabla \boldsymbol{u} =\rho \boldsymbol{f} - \nabla p +\nabla(\eta \vartheta)+\nabla \cdot(2 \mu \boldsymbol{D}) \label{eq:Momentum_balance} \\ 
& \text{kinetic energy equation:}\quad &\rho \boldsymbol{u} \cdot \nabla \left(\frac{1}{2} |\boldsymbol{u}|^{2}\right) =\rho \boldsymbol{f} \cdot \boldsymbol{u}+p \vartheta+\nabla \cdot(\boldsymbol{T} \cdot \boldsymbol{u})-\phi \label{eq:kinetic_energy_balance}
\end{align}
where $\rho$, $\boldsymbol{u}$, $\boldsymbol{f}$, $p$, $\eta$, $\mu$, and $\boldsymbol{q}$ are the density, velocity, body force, pressure, dilatation viscosity, shear viscosity, and heat flux of the flow, respectively. $\vartheta=\nabla \cdot \boldsymbol{u}$, $\boldsymbol{D}=\left[\nabla \boldsymbol{u}+(\nabla \boldsymbol{u})^{T}\right] / 2$, and $\mathbf{T}=(-p+\eta \vartheta) \mathbf{I}+2 \mu \mathbf{D}$ are the velocity divergence, strain-rate tensor, and stress tensor, respectively. $\phi$ is the dissipation of kinetic energy:
\begin{equation}
\phi=\eta \vartheta^{2}+2 \mu \boldsymbol{D} : \boldsymbol{D} \label{eq:dissipation}
\end{equation}
\par Helmholtz and Rayleigh (\cite{helmholtz1868theorie,rayleigh1913lxv,serrin1959mathematical}) proved that, for an incompressible viscous fluid, if the acceleration $\boldsymbol{a}=\boldsymbol{u} \cdot \nabla \boldsymbol{u}$ can be derived by a potential $\zeta$ ($\boldsymbol{a}=\nabla \zeta$ or $\nabla \times \boldsymbol{a} = 0$), it should possess minimal dissipation, which is the well-known Helmholtz-Rayleigh minimal dissipation theorem. \cite{He1988} generalized this theorem to compressible flows (\cite{wu2007vorticity}). Here, we demonstrate the proof process concisely, and provide conditions that compression flows should satisfy. The total dissipation $\Phi$ in a control volume $V$ bounded by $\ell$ is considered, where $V$ is nondeformable or the flow on $\ell$ (if $V$ is deformable) is nondissipative. With the constraint provided by Equation (\ref{eq:Mass_conservation}), the variation of $\Phi$ can be written as
\begin{equation}
\delta \Phi=\delta \int_{V}[\phi+\lambda \nabla \cdot(\rho \boldsymbol{u})] d V=0
\end{equation}
where $\lambda$ is a Lagrangian multiplier and $\mathcal{L}=\phi+\lambda \nabla \cdot(\rho \boldsymbol{u})$ is the Lagrangian. Because $\boldsymbol{u}$ and $\rho$ are the two independent variables of $\mathcal{L}$, the Eular-Lagrangian equations are
\begin{align}
& \frac{\delta \mathcal{L}}{\delta \boldsymbol{u}}=0 :\quad \frac{\partial \mathcal{L}}{\partial \boldsymbol{u}}-\nabla \cdot \frac{\partial \mathcal{L}}{\partial \nabla \boldsymbol{u}}-\nabla \frac{\partial \mathcal{L}}{\partial \nabla \cdot \boldsymbol{u}}=0 \Rightarrow 2[\nabla(\eta \vartheta)+\nabla \cdot(2 \mu \boldsymbol{D})]+ \rho \nabla \lambda=0  \label{eq: varation u} \\
& \frac{\delta \mathcal{L}}{\delta \rho}=0 :\quad \frac{\partial \mathcal{L}}{\partial \rho}-\nabla \cdot \frac{\partial \mathcal{L}}{\partial \nabla \rho}=0 \quad \Rightarrow \quad \boldsymbol{u} \cdot \nabla \lambda=0 \label{eq: varation rho} 
\end{align}
If a flow satisfies (i) $\boldsymbol{a}=\nabla \zeta$; (ii) $\boldsymbol{f} = - \nabla U$, i.e., the body force can be derived by a potential $U$; (iii) $\nabla p / \rho = \nabla \int d p / \rho$ or $\nabla p \times \nabla \rho=0$, i.e., the flow is barotropic, the viscous force can then be derived by a potential $\xi$ from function (\ref{eq:Momentum_balance}), i.e., $[\nabla(\eta \vartheta) + \nabla \cdot(2 \mu D)] / \rho = \nabla \xi$. Thus, with $\lambda$ chosen as $\lambda=-\left(\zeta+\int d p / \rho + U + \xi \right)$, Equations (\ref{eq: varation u}) and (\ref{eq: varation rho}) can be exactly rearranged to the momentum equation (\ref{eq:Momentum_balance}) and kinetic energy equation (\ref{eq:kinetic_energy_balance}), respectively. Therefore, compressible flows satisfying (i), (ii), and (iii) must have minimal dissipation.
\par For a flow passing through a straight shock wave, the acceleration $\boldsymbol{a}$ can be decomposed into two parts relative to the shock front, i.e., the vertical component $a_{n}$ and tangential one $a_{\tau}$ (consider two-dimensional cases). Because $\boldsymbol{u}$ only changes perpendicularly through the shock, there must be $\partial a_{n} / \partial \tau = 0$ and $a_{\tau} = 0$, then $\left|\nabla \times \boldsymbol{a}\right| = \partial a_{\tau} / \partial n - \partial a_{n} / \partial \tau = 0$, satisfying condition (i). Condition (ii) is also satisfied because the body force $\boldsymbol{f}$ is gravity, which can be negligible. Note that $\nabla p$ and $\nabla \rho$ are both perpendicular to the shock front, i.e., $\nabla p \times \nabla \rho=0$; thus, condition (iii) is satisfied. Therefore, a steady flow across a straight shock wave has minimal dissipation. A corollary of this demonstration is that, if the total dissipation of a steady flow is only contributed by shock waves, this flow should have minimal dissipation. A piece of evidence is that, although two (one weak and one strong) oblique shock waves are both theoretically possible for the same deflection angle, the observable shock wave, in practice, is always the weak one.
\section{Dissipation of the flow system}
\par In this section, we will demonstrate that, for a large separation, the total dissipation $\Phi$ of this flow system is primarily contributed by shock waves, and show the dependency of $\Phi$ on $\theta_{s}$.
\par As \ref{fig:compression-ramp-information} (b) shows, four types of flow structure probably generate dissipation, i.e., (i) shock waves, (ii) shear layers, (iii) the separation bubble, and (iv) the expansion fan (behind the triple point $B$). For (i), a shock wave with length $L_{\omega}$, the dissipation can be estimated as $\int_{L_{\omega}} \int_{\epsilon_{\omega}} \phi d n d \tau \sim \mathcal{O}(\mu_{\omega} \Delta u_{\omega}^{2} L_{\omega} / \epsilon_{\omega})$ using Equation (\ref{eq:dissipation}), where $L_{\omega}$, $\epsilon_{\omega}$, $\mu_{\omega}$, and $\Delta u_{\omega}$ are the characteristic length, thickness, viscosity, and velocity difference of this shock, respectively. For (ii), a shear layer with length $L_{\delta}$, its dissipation can also be estimated, i.e., $\int_{L_{\delta}} \int_{\epsilon_{\delta}} \phi d n d\tau \sim \mathcal{O}(\mu_{\delta} \Delta u_{\delta}^{2} L_{\delta} / \epsilon_{\delta})$, where $\epsilon_{\delta}$, $\mu_{\delta}$, and $\Delta u_{\delta}$ are the characteristic thickness, viscosity, and velocity difference of this shear layer, respectively. Because $\mu_{\delta} \sim \mu_{\omega}$, $\Delta u_{\delta} \sim \Delta u_{\omega}$, $L_{\delta} \sim L_{\omega}$, and $\epsilon_{\delta} \gg \epsilon_{\omega}$, dissipations induced by shear layers are negligible relative to those induced by shock waves. For (iii), a separation bubble $\Omega_{s}$ with characteristic length $L_{s}$ (i.e., $L_{s}^2 \sim \mathcal{O} (\Omega_{s})$), its dissipation can be estimated as $\int_{\Omega_{s}} \phi d V \sim \mathcal{O} (\mu_{s} \Delta u_{s}^2)$, where $\mu_{s}$ is the characteristic viscosity of the bubble. Although its internal vortex structure may be complex, its characteristic velocity $\Delta u_{s}$ is always very small relative to $\Delta u_{\omega}$, i.e., $\Delta u_{s} \ll \Delta u_{\omega}$. In consideration of $L_{\omega} / \epsilon_{\omega} \gg 1$, the dissipation induced by the separation bubble can be negligible as well. For (iv), an expansion fan with area $\Omega_{f}$, its dissipation $\int_{\Omega_{f}} \phi d V$ can be calculated using the relation $\phi=\rho T d s / d t - \nabla \cdot \boldsymbol{q}$, i.e., the deformation of the energy equation (\cite{wu2007vorticity}), where $d s / d t$, $T$, and $\boldsymbol{q}$ are the entropy production, temperature, and heat conductivity of the flow, respectively. Because the flow passing through $\Omega_{f}$ is an isentropic process, and the heat flux $\boldsymbol{q}$ on the boundary $\partial \Omega_{f} $ of $\Omega_{f}$ is zero, there must be $\int_{\Omega_{f}} (\rho T d s / d t - \nabla \cdot \boldsymbol{q}) d V = 0$. Therefore, the dissipation induced by the expansion fan is zero. In this flow system, the appropriate control volume $V$ is chosen and is bounded by $AB$, $BG$, $GC$, $CS$ and $SA$, as shown in Figure \ref{fig:compression-ramp-information} (b). Because the total dissipation in $V$ is primarily contributed by shock waves $SB$ and $RB$, a steady flow in $V$ should have minimal dissipation, which implies $\theta_{s}$ of a steady state should cause shock waves $SB$ and $RB$ to dissipate minimal kinetic energy.
\par Subsequently, we will illustrate how the total dissipation $\Phi$ depends on $\theta_{s}$. By integrating function (\ref{eq:kinetic_energy_balance}) across the shock wave perpendicularly, we obtain the dissipation induced by a shock wave per unit length:
\begin{equation}
\begin{aligned}
\widehat{\mathcal{\phi}}& = \int_{\epsilon_{\omega}} \phi d n = \mathcal{E}-\mathcal{P} \\
\mathcal{E} & = - \int_{\epsilon_{\omega}} \rho \boldsymbol{u} \cdot \nabla \left(\frac{1}{2} |\boldsymbol{u}|^{2}\right) d n = \frac{1}{2}\left\{\rho_{a}\left(M_{a} \mathit{c}_{a} \sin \beta\right)^{3} - \rho_{b}\left[M_{b} \mathit{c}_{b} \sin \left(\beta-\theta\right)\right]^{3}\right\}\\
\mathcal{P} & = - \int_{\epsilon_{\omega}} p \vartheta d n \approx \frac{1}{2} \left(p_{a} + p_{b}\right) \left[M_{a} \mathit{c}_{a} \sin \beta - M_{b} \mathit{c}_{b} \sin \left(\beta-\theta\right) \right]\\
\label{eq:dissipation_per_unit_length}
\end{aligned}
\end{equation}
where $\mathcal{E}$ and $\mathcal{P}$ are the kinetic energy loss and the negative work of pressure, respectively, implying that one portion of the kinetic energy loss is stored as the potential energy and the other is dissipated. In Equation (\ref{eq:dissipation_per_unit_length}), $M$, $\mathit{c}$, and $\beta$ are the Mach number, acoustic velocity, and shock angle, respectively. The subscripts `a' and `b' are the locations ahead of and behind a shock wave, respectively. As \ref{fig:compression-ramp-information} (b)  shows, `a' and `b' correspond to `0' and `1' for shock $SB$, and to `1' and `2' for $RB$, respectively. Quantities in Equation (\ref{eq:dissipation_per_unit_length}) satisfy the following:
\begin{equation}
\left.\begin{array}{rl}
&{M_{1}^{2}=\mathcal{F}_{M}\left(M_{0}, \beta_{s}\right),\quad M_{2}^{2}=\mathcal{F}_{M}\left(M_{1}, \beta_{r}\right)} \\
&{\mathit{c}_{0} / \mathit{c}_{1}=\mathcal{F}_{\mathit{c}}\left(M_{0}, \beta_{s}\right),\quad \mathit{c}_{1} / \mathit{c}_{2}=\mathcal{F}_{\mathit{c}}\left(M_{1}, \beta_{r}\right)} \\
&{\rho_{0} / \rho_{1}=\mathcal{F}_{\rho}\left(M_{0}, \beta_{s}\right),\quad \rho_{1} / \rho_{2}=\mathcal{F}_{\rho}\left(M_{1}, \beta_{r}\right)} \\
&{p_{0} / p_{1}=\mathcal{F}_{p}\left(M_{0}, \beta_{s}\right),\quad p_{1} / p_{2}=\mathcal{F}_{p}\left(M_{1}, \beta_{r}\right)} \\
&{\mathcal{F}_{\beta}\left(M_{0}, \beta_{s}, \theta_{s}\right)=0,\quad \mathcal{F}_{\beta}\left(M_{1}, \beta_{r}, \theta_{r}\right)=0} \\
&{\theta_{r}=\alpha-\theta_{s}}
\end{array}\right\}
\end{equation}
where $\beta_{s}$ and $\beta_{r}$ are the shock angles of $SB$ and $RB$, respectively, and $\theta_{r}$ is the deflection angle of the flow across $RB$. $\mathcal{F}_{M}$, $\mathcal{F}_{\mathit{c}}$, $\mathcal{F}_{\rho}$, $\mathcal{F}_{p}$, and $\mathcal{F}_{\beta}$ are the classical Rankine-Hugoniot relations (\cite{rankine1870xv,rankine1887propagation}):
\begin{equation}
\begin{aligned}
& \mathcal{F}_{M}(M, \beta) \equiv \frac{M^{2}+\frac{2}{\gamma-1}}{\frac{2 \gamma}{\gamma-1} M^{2} \sin ^{2} \beta-1}+\frac{M^{2} \cos ^{2} \beta}{\frac{\gamma-1}{2} M^{2} \sin ^{2} \beta+1}\\
& \mathcal{F}_{\mathit{c}}(M, \beta) \equiv \frac{\left[(\gamma-1) M^{2} \sin ^{2} \beta+2\right]^{1 / 2}\left[2 \gamma M^{2} \sin ^{2} \beta-(\gamma-1)\right]^{1 / 2}}{(\gamma+1) M \sin \beta}\\
& \mathcal{F}_{\rho}(M, \beta) \equiv \frac{(\gamma+1) M^{2} \sin ^{2} \beta}{(\gamma-1) M^{2} \sin ^{2} \beta+2},\quad \mathcal{F}_{p}(M, \beta) \equiv \frac{2 \gamma}{\gamma+1} M^{2} \sin ^{2} \beta-\frac{\gamma-1}{\gamma+1} \\
& \mathcal{F}_{\beta}(M, \beta, \theta) \equiv 2 \cot \beta \frac{M^{2} \sin ^{2} \beta-1}{M^{2}(\gamma+\cos 2 \beta)+2}-\tan \theta
\label{eq: R_H_relation}
\end{aligned}
\end{equation}
Therefore, for a given inflow Mach number $M_{0}$ and ramp angle $\alpha$, both $\widehat{\mathcal{\phi}}_{ST}$ and $\widehat{\mathcal{\phi}}_{RT}$ depend only on the order parameter $\theta_{s}$. We will now show the geometrical relationships of the flow system. For a large separation, the size (area) of the separation bubble $\Omega_{s}$ can be calculated using the triangle area formula:
\begin{equation}
\Omega_{s} = \frac{L_{SR}^{2}\sin\theta_{r}\sin\theta_{s}}{2\sin\alpha}
\label{eq:Omega_s}
\end{equation}
where $L_{SR}$ is the length of the free shear layer starting from the separation point $S$ to the reattachment point $R$. Note that the flow at a low velocity in the separation bubble is approximately incompressible, where the pressure, temperature, and density are $p_{1} (\theta_{s})$, $T_{w}$, and $\rho_{s} (\theta_{s})=\gamma M_{0}^{2} p_{1}(\theta_{s}) / T_{w a l l}$, respectively. Considering that a stronger adverse pressure gradient can press more flows into the separation bubble, we assume that $\Pi_{s}(\theta_{s}) \propto p_{1}(\theta_{s}) / p_{0}$ as $\theta_{s}$ varies, where $\Pi_{s}(\theta_{s}) = \rho_{s}(\theta_{s}) \Omega_{s}$ is the mass of the flow in the bubble. Thus, $\Omega_{s} \propto T_{\text {wall}} /\left(\rho_{0} T_{0}\right)$ remains constant as $\theta_{s}$ varies. By normalizing formula (\ref{eq:Omega_s}) with $\Omega_{s}$, we obtain
\begin{equation}
\ell_{SR} =\sqrt{\frac{2\sin\alpha}{\sin\theta_{r}\sin\theta_{s}}}
\label{eq:dimensionless_length}
\end{equation}
where $\ell_{SR}$ is the dimensionless length of $L_{SR}$ being nondimensionalized using $\sqrt{\Omega_{s}}$. The coordinate of point $S$ is set as $(0,0)$; thus points $B$ and $R$ are
\begin{equation}
\begin{aligned}
&x_{B} = \ell_{SR} \frac{\sin\theta_{s}-\cos\theta_{s} \tan(\beta_{r}+\theta_{s})}{\tan\theta_{s}-\tan(\beta_{r}+\theta_{s})}, y_{B} = x_{B}\tan\beta_{s}\\
&x_{R} = \ell_{SR}\cos\theta_{s}, y_{R} = \ell_{SR}\sin\theta_{s}
\end{aligned}
\label{eq:point_TR}
\end{equation}
Thus, the lengths of shock wave $SB$ and $RB$ can be obtained:
\begin{equation}
\begin{aligned}
&\ell_{SB} = \sqrt{x_{B}^2 + y_{B}^2}\\
&\ell_{RB} = \sqrt{(x_{B}-x_{R})^2 + (y_{B}-y_{R})^2}
\end{aligned}
\label{eq:length_of_ST_RT}
\end{equation}
Using Equations (\ref{eq:dissipation_per_unit_length}) and (\ref{eq:length_of_ST_RT}), we can calculate the total dissipation $\Phi$ in $V$:
\begin{equation}
\Phi(\theta_{s})=\widehat{\phi}_{S B} \ell_{S B}+\widehat{\phi}_{R B} \ell_{R B}
\label{eq:total_dissipation}
\end{equation}
\section{Results}
\begin{figure*}
	\begin{center}
		\subfigure[\label{subfig:Figure4a}]{\includegraphics[width = 0.31\linewidth]{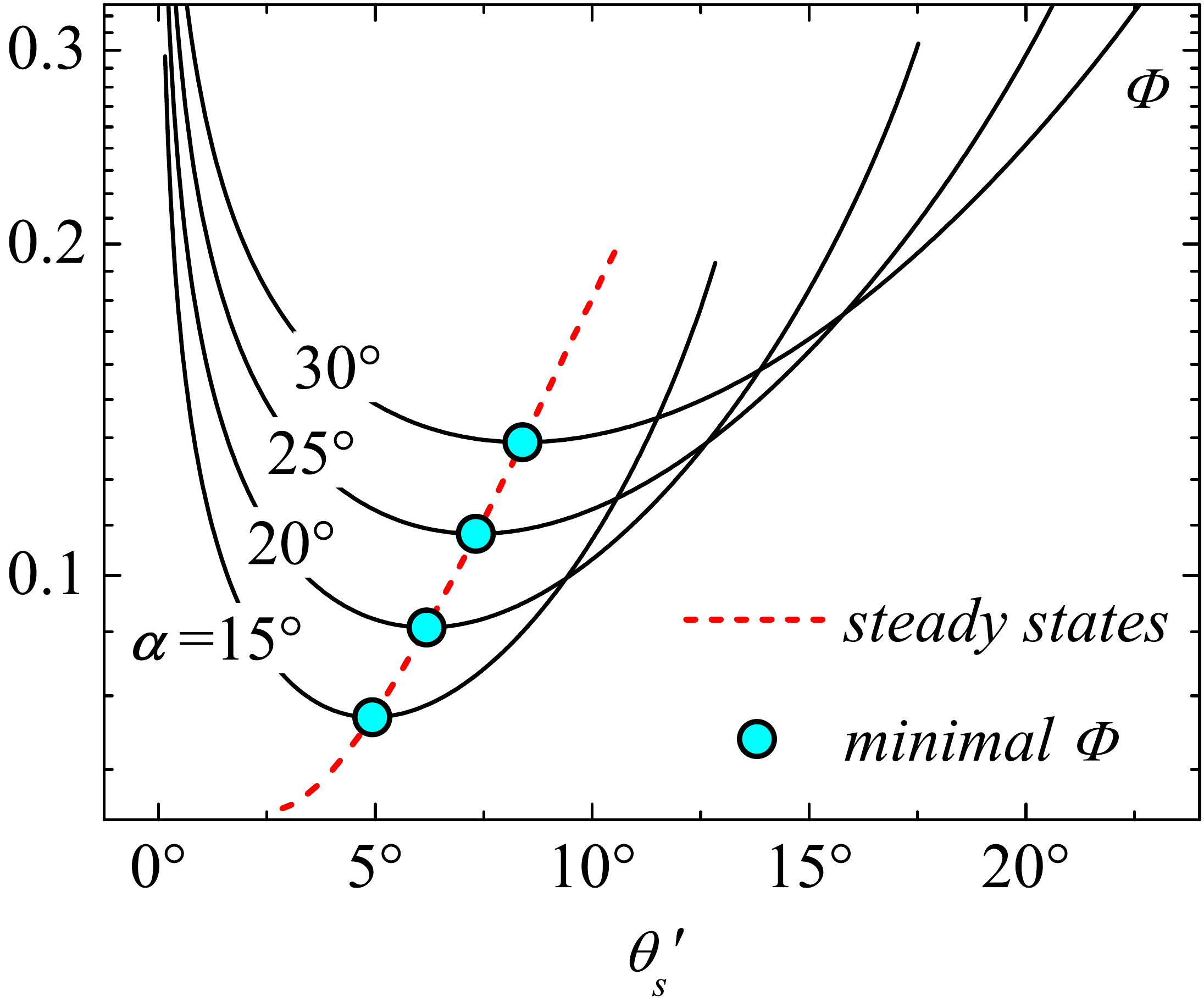}}
		\subfigure[\label{subfig:Figure4b}]{\includegraphics[width = 0.31\linewidth]{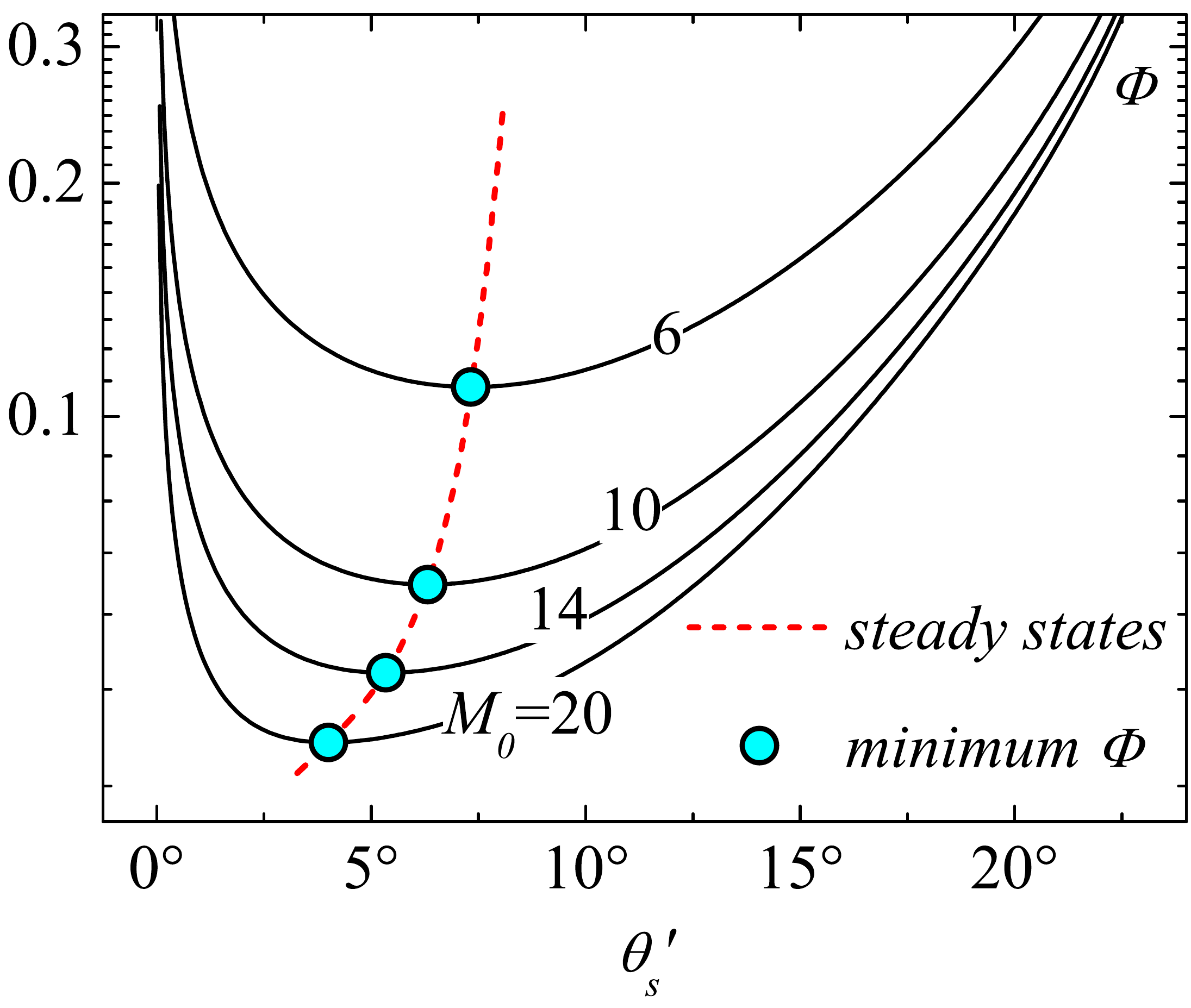}}
		\subfigure[\label{subfig:theta_VS_angle}]{\includegraphics[width = 0.32\linewidth]{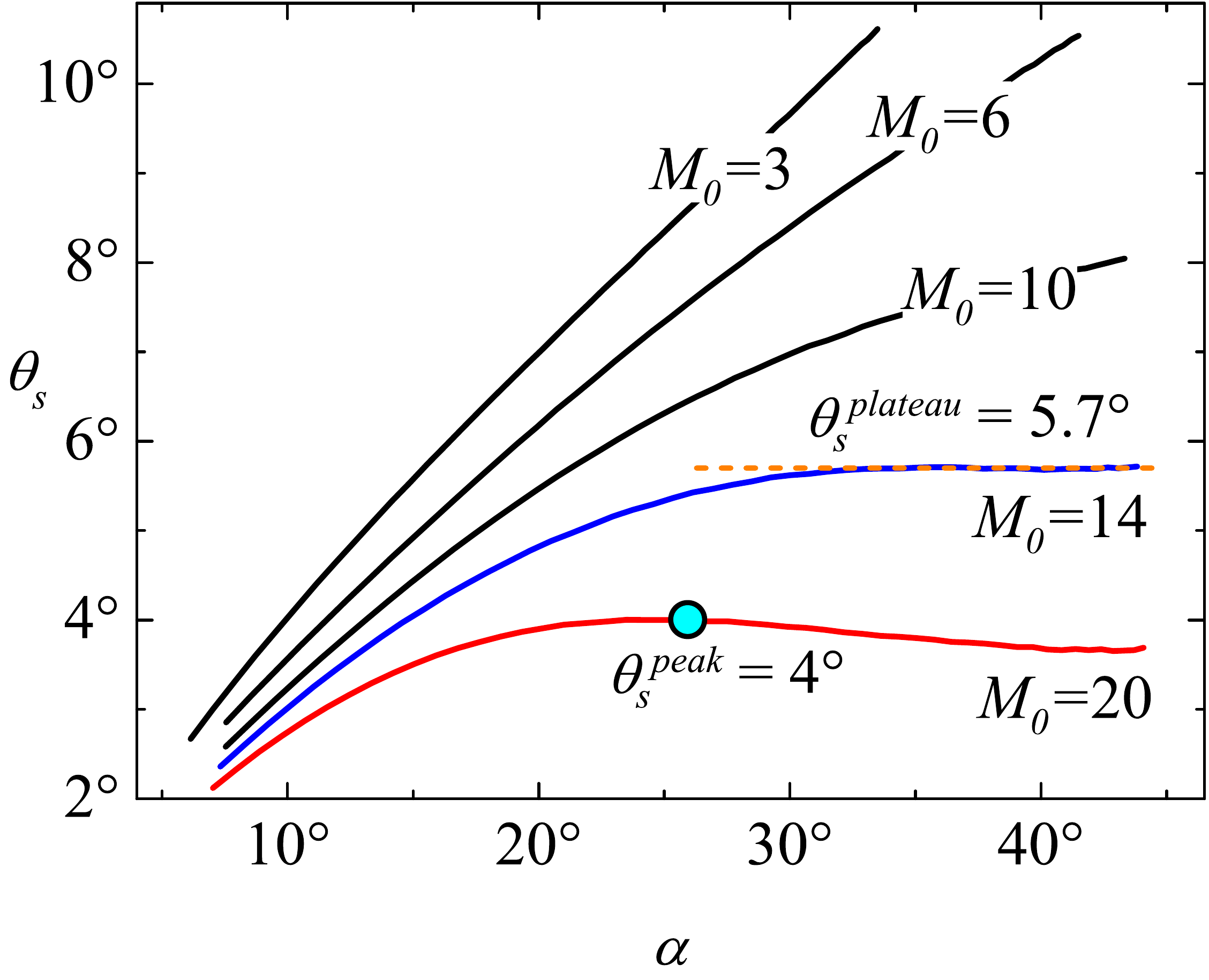}}
	\end{center}
	\caption{Theoretical $\theta_{s}$ corresponding to minimal dissipation. (a) and (b) depict $\Phi$ at all possible $\theta_{s}^{\prime}$, where the cyan points are the minimal dissipation, and the red dash lines are the theoretical prediction results. (a) $\alpha = 15, 20, 25$, and $30\degree$ with $M_{0} = 6$; (b) $M_{0} = 20, 14, 10$, and $6$ at $\alpha = 25\degree$. (c) Theoretical $\theta_{s}$ varying with $\alpha$ at different $M_{0}$. \label{fig:minimum_dissipation}}
\end{figure*}
\par Figure \ref{fig:minimum_dissipation} (a) and (b) depict the variation of the total dissipation $\Phi$ with all possible separation angle $\theta_{s}^{\prime}$, where (a) depicts $\Phi$ at different ramp angles $\alpha$ with inflow Mach number $M_{0} = 6$, and (b) depicts $\Phi$ with different $M_{0}$ at $\alpha = 25\degree$. As demonstrated in \textbf{Sections 2} and \textbf{3}, steady states of the flow system must have minimal dissipation. Therefore, if a steady state is at $\theta_{s}$, there must be
\begin{equation}
\frac{\partial \Phi}{\partial \theta_{s}^{\prime}}\bigg|_{\theta_{s}}=0,\quad \frac{\partial^2 \Phi}{\partial \theta_{s}^{\prime 2}}\bigg|_{\theta_{s}}>0 
\label{eq: local_min_differential_relation}
\end{equation}
Only one minimal value of $\Phi$ can be observed to exist for a given $M_{0}$ and $\alpha$, indicating that only one steady state exists in a compression ramp with large separation, which corresponds with all experimental observations. Figure \ref{subfig:theta_VS_angle} shows the variation in the theoretical $\theta_{s}$ with different $M_{0}$ and $\alpha$. For a given $\alpha$, $\theta_{s}$ is observed to decrease as $M_{0}$ increases. However, for a given $M_{0}$, three possible scenarios occur, i.e., (i) $\theta_{s}$ increases with $\alpha$ ($M_{0} = 3, 6, 10$); (ii) $\theta_{s}$ increases when $\alpha$ is moderate but maintains at a plateau when $\alpha$ is sufficiently large (for $M_{0} = 14$, $\theta_{s} \approx 5.7\degree$ when $\alpha > 35\degree$); (iii) there is a critical $\alpha$ corresponding to a peak $\theta_{s}$ (for $M_{0} = 20$, critical $\alpha \approx 26\degree$ and $\theta_{s}^{peak} \approx 4\degree$), over which $\theta_{s}$ will decrease as $\alpha$ increases.
\begin{figure*}
	\begin{center}
		\subfigure[\label{subfig:field_comparison_a}]{\includegraphics[width = 0.49\linewidth]{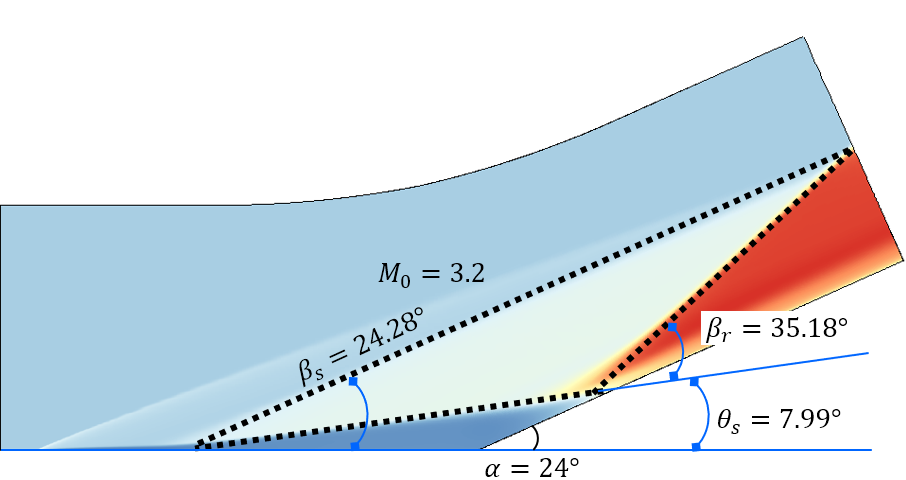}}
		\subfigure[\label{subfig:field_comparison_b}]{\includegraphics[width =  0.49\linewidth]{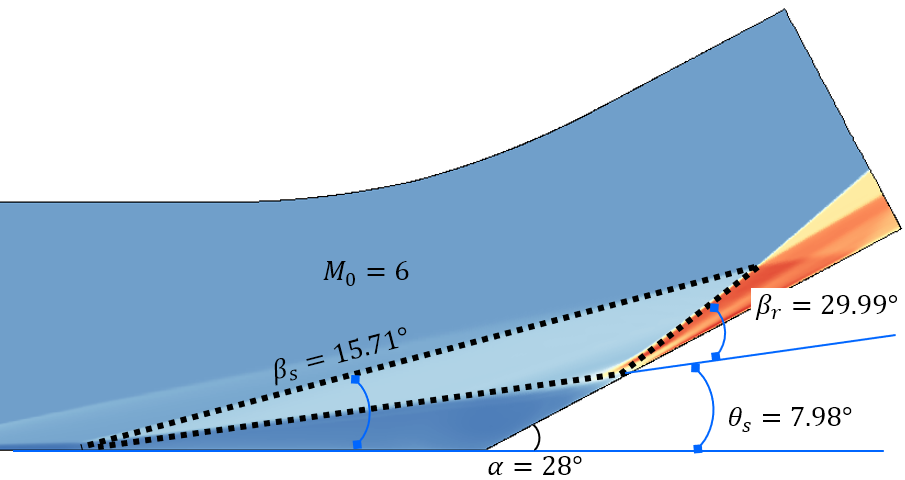}}
		\subfigure[\label{subfig:field_comparison_c}]{\includegraphics[width = 0.49\linewidth]{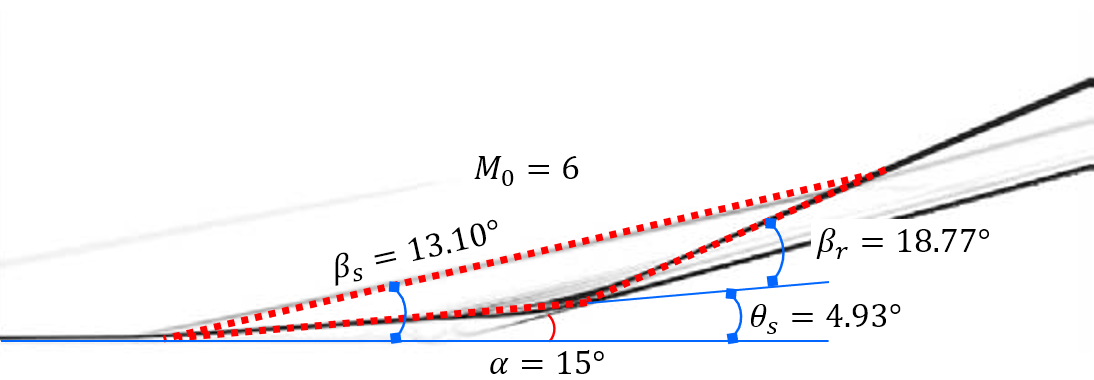}}
		\subfigure[\label{subfig:field_comparison_d}]{\includegraphics[width =  0.49\linewidth]{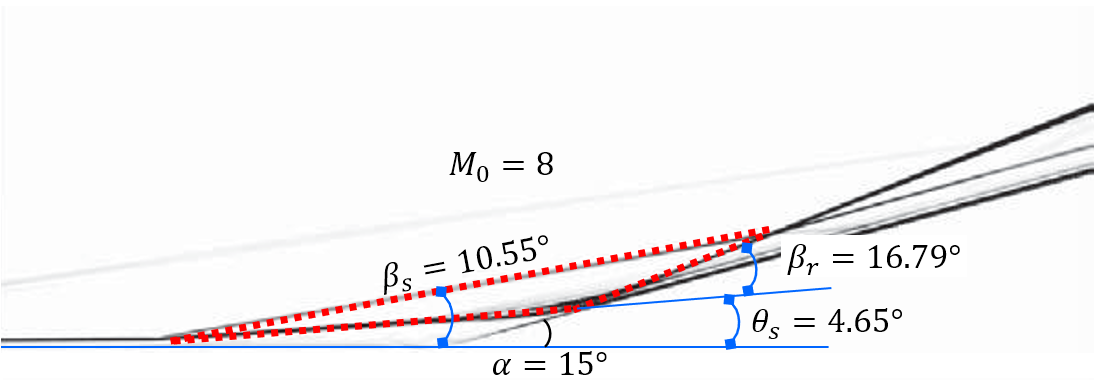}}
		\subfigure[\label{subfig:field_comparison_e}]{\includegraphics[width =  0.49\linewidth]{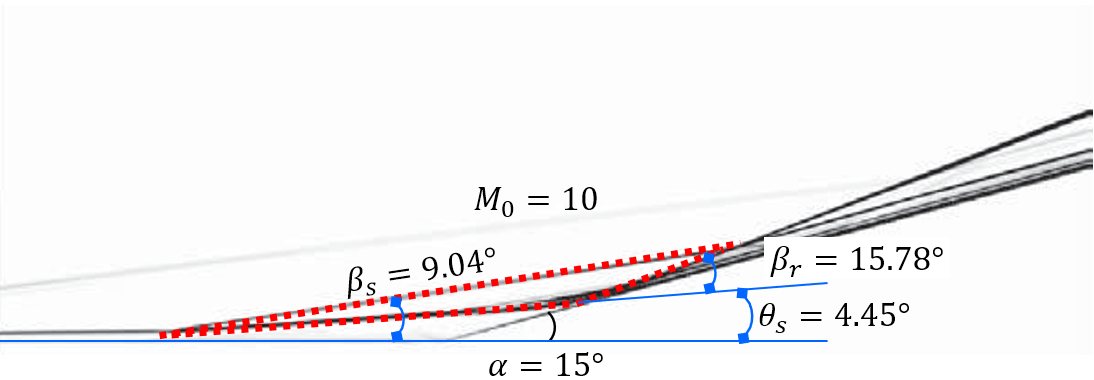}}
		\subfigure[\label{subfig:field_comparison_f}]{\includegraphics[width =  0.49\linewidth]{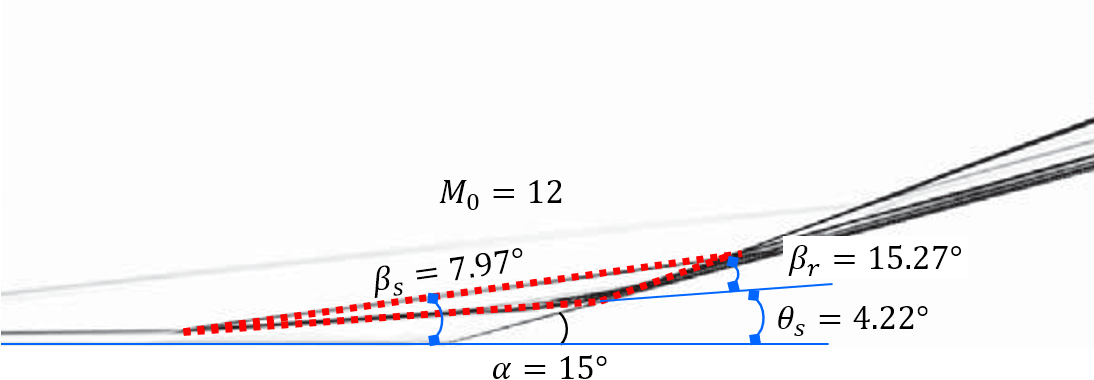}}
		\subfigure[\label{subfig:field_comparison_g}]{\includegraphics[width =  0.49\linewidth]{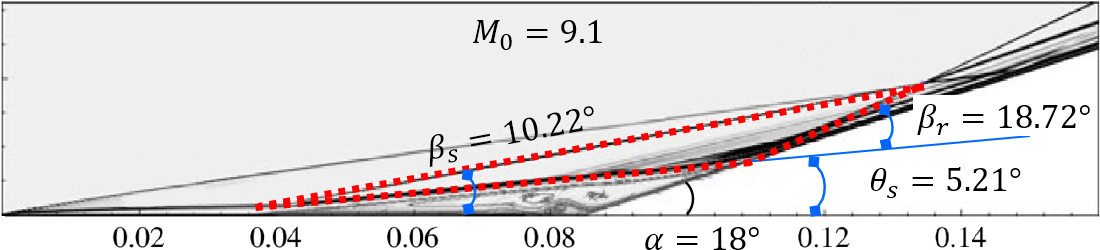}}
		\subfigure[\label{subfig:field_comparison_h}]{\includegraphics[width =  0.49\linewidth]{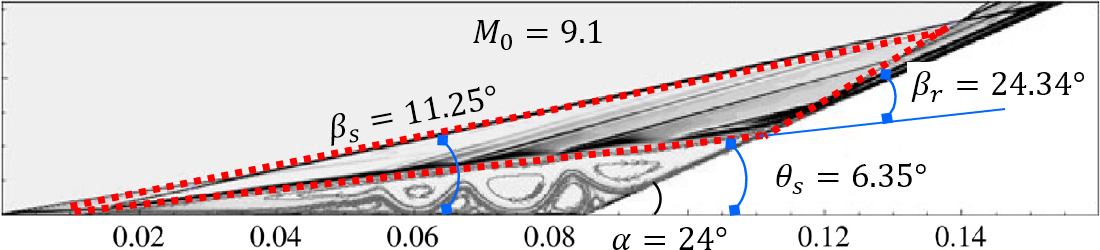}}
	\end{center}
	\caption{Comparison of the proposed theoretical $\theta_{s}$, $\beta_{s}$ and $\beta_{r}$ (dash line) with numerical results. We simulated (a) $M_{0} = 3.2$ and $\alpha = 24\degree$ and (b) $M_{0} = 6$ and $\alpha = 28\degree$ and coloured them by density contours; (c)--(f) are extracted from \citet[pp. 318-319]{babinsky2011shock} at $\alpha = 15 \degree$, where (c) $M_{0} = 6$, (d) $M_{0} = 8$, (e) $M_{0} = 10$, and (f) $M_{0} = 12$; (g) and (h) are extracted from \cite{gai2019hypersonic} with $M_{0} = 9.1$, where (g) $\alpha = 18\degree$ and (h) $\alpha = 24\degree$. \label{fig:field_comparison}}
\end{figure*}
\par
\begin{figure}
	\centerline{\includegraphics[width = 0.8\columnwidth]{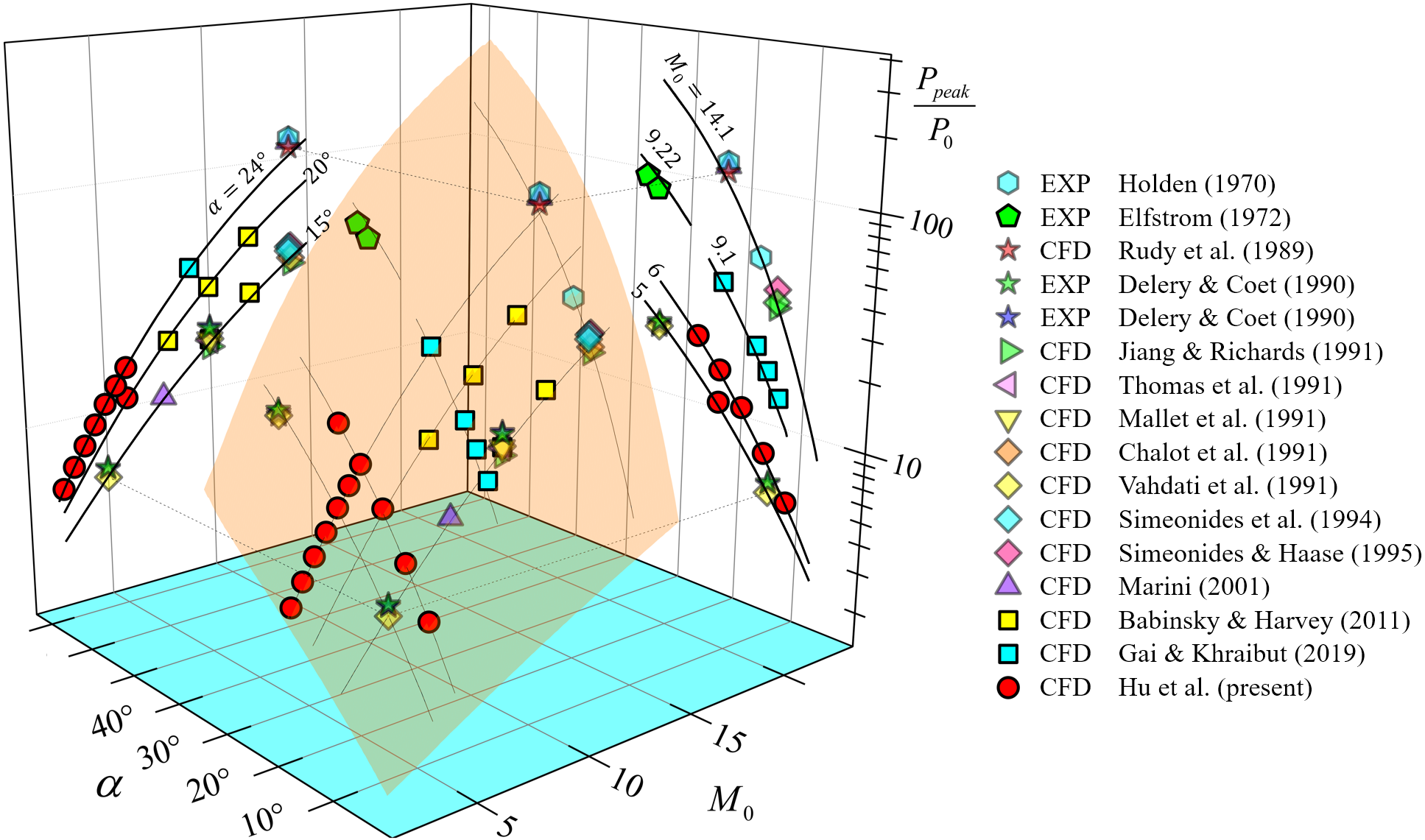}} 
	\caption{Comparison of theoretical pressure peak $p_{peak}$ with numerical and experimental results. The three-dimensional orange surface is the solution of $p_{peak}/p_{0}$ depending on both $M_{0}$ and $\alpha$. The thick black lines on the ($M_{0}$,$p_{peak}/p_{0}$)-plane and ($\alpha$,$p_{peak}/p_{0}$)-plane are projections of the thin ones on the theoretical $p_{peak}/p_{0}$-surface.}
	\label{fig:3D_surface}
\end{figure}

  Figure \ref{fig:field_comparison} shows the comparison of the proposed theoretical $\theta_{s}$, $\beta_{s}$, and $\beta_{r}$ with numerical results from different researchers (our DNSs, \citet[pp. 318-319]{babinsky2011shock} and \cite{gai2019hypersonic}). The theory in this paper can be observed to agree well with the flow patterns encompassing a wide range of Mach numbers and ramp angles ($M_{0}$ varying from 3.2 to 12, $\alpha$ varying from $15$ to $28\degree$). As $M_{0}$ and $\alpha$ increase, the pressure peak $p_{peak}$ behind the reattachment shock wave, located at subsystem 2 in Figure \ref{fig:compression-ramp-information} (b), increases rapidly and will be two orders of magnitude larger than the free-stream pressure $p_{0}$. Figure \ref{fig:3D_surface} depicts the comparison of theoretical $p_{peak}$ with that of previous works, including experimental (\citep{holden1970theoretical,elfstrom1972turbulent,delery1990experimental,delery1991experiments}) and numerical (\cite{rudy1989validation,jiang1991hypersonic,thomas1991grid,vahdati1991application,mallet1991contribution,chalot1991application,simeonides1994experimental,simeonides1995experimental,marini2001analysis,babinsky2011shock,gai2019hypersonic}) results. The theoretical $p_{peak}/p_{0}$ is observed to agree well with all the numerical and experimental results with a $M_{0}$ varying from 3.2 to 14.1 and $\alpha$ varying from $15$ to $38\degree$. Furthermore, for a large separation, as Table \ref{tab:five-conditions-table} shows, the theoretical predictions are independent of the the Reynolds number $Re$ and the wall temperature $T_{w}$, which is because the dissipation induced by shear layers can be negligible relative to shock waves, as demonstrated in \textbf{Section 3}. Additionally, as the maximum heat flux $h_{peak}$ generated in the reattachment region can be correlated with $p_{peak}$ in terms of simple power-law relations (\cite{holden1978study}), the proposed theory can be used to estimate the $h_{peak}$ quantificationally, which is vital in both supersonic and hypersonic flows.
\begin{table}
	\begin{tabular}{@{}cccccccccc@{}}
		\toprule
		\multirow{2}{*}{$M_{0}$} & \multirow{2}{*}{$\alpha$} & \multirow{2}{*}{\begin{tabular}[c]{@{}c@{}}$Re_{0}$\\ ($m^{-1}$)\end{tabular}} & \multirow{2}{*}{\begin{tabular}[c]{@{}c@{}}$T_{w}$\\ ($K$)\end{tabular}} & \multirow{2}{*}{\begin{tabular}[c]{@{}c@{}}$T_{0}$\\ ($K$)\end{tabular}} & \multirow{2}{*}{state} & \multicolumn{2}{c}{$p_{peak}/p_{0}$} & $p_{3} / p_{0}$ & \multicolumn{1}{c}{\multirow{2}{*}{author}} \\ \cmidrule(lr){7-9}
		&  &  &  &  &  & theory & EXP/CFD & inviscid & \multicolumn{1}{c}{} \\ \cmidrule(r){1-6} \cmidrule(l){10-10} 
		14.1 & 24\degree & $2.362 \times 10^5$ & 297.22 & 88.88 & L & 102.95 & 123.5 (EXP) & 58.02 & \multirow{2}{*}{Holden} \\
		14.1 & 18\degree & $2.362 \times 10^5$ & 297.22 & 88.88 & L & 54.52 & 51.88 (EXP)& 34.23 &  \\
		9.22 & 38\degree & $4 \times 10^5 (Re_{L_{\delta}})$ & 295 & 59.44 & T & 106.08 & 93.13 (EXP) & 58.28 & Elfstrom \\
		9.1 & 24\degree & $3.22 \times 10^6$ & 351.24 & 160 & L & 36.3 & 37.32 (CFD) & 25.48 & Gai \& Khraibut \\
		6.0 & 28\degree & $2.79 \times 10^6$ & 307 & 108.1 & L & 20.04 & 20.15 (CFD) & 15.84 & Hu et al. \\ \cmidrule(r){1-6} \cmidrule(l){10-10} 
	\end{tabular}
	\caption{Comparison of theoretical pressure peak $p_{peak}$ with numerical and experimental results for five conditions encompassing different Reynolds numbers and wall temperature. $Re_{L_{\delta}}$ is the Reynolds number of the boundary layer in the upstream of the separation point, of which the characteristic thickness is $L_{\delta}$, and the states `L' and `T' are laminar and turbulent, respectively. $p_{3} / p_{0}$ is the pressure rise of the flows through the inviscid shock wave.}
	\label{tab:five-conditions-table}
\end{table}
\section{Conclusion}
\par In this work, the least action principle is used to reveal that the synergic principle of a compression ramp flow with large separation is the minimal dissipation theorem. Based on this theorem, we obtain theoretical flow patterns and the pressure peak $p_{peak}$, which agree very well with the numerical and experimental results for a wide range of Mach numbers $M_{0}$ and ramp angles $\alpha$. Because the maximum heat flux $h_{peak}$ at the reattachment region can be correlated with $p_{peak}$ in terms of simple power-laws, $h_{peak}$ can be estimated quantificationally, which is vital in aerospace engineering. Furthermore, the predicted results are independent of the Reynolds number $Re$ and wall temperature $T_{w}$. The present theoretical framework has both a strict mathematical logic and a clear physical image, which are expected to be applied to other flow systems dominated by shock waves.

\section*{Acknowledgement}
We are grateful to Professor You-Sheng Zhang for his helpful discussion and to Professor Xin-Liang Li for his support of the numerical simulation. This work is supported by National Key R \& D Program of China (Grant No.2019YFA0405300).
\bibliographystyle{jfm}
\bibliography{jfm-instructions}

\end{document}